\documentclass[%
 preprint, %linenumbers,
superscriptaddress,
 amsmath,amssymb,
 aps, 
prapplied,
showkeys
]{revtex4-2}

\usepackage{graphicx}% Include figure files
\usepackage{dcolumn}% Align table columns on decimal point
\usepackage{bm}% bold math
\usepackage[bookmarks=true,unicode=true,linktocpage=true]{hyperref}
\hypersetup{colorlinks=true, allcolors=blue}  

\begin{document}

% \preprint{APS/123-QED}

\title{\textbf{Efficient Lasing in MoS\textsubscript{2}/WSe\textsubscript{2}-Based Metasurfaces Enabled by Quasi-Dark Magnetic Dipole Resonance}}

\author{Georgios Nousios}
\email{gnnousios@ece.auth.gr}
    \affiliation{School of Electrical and Computer Engineering, Aristotle University of Thessaloniki (AUTH), GR-54124 Thessaloniki, Greece}
\author{Thomas Christopoulos} 
    \affiliation{Laboratoire Photonique, Num\'erique et Nanosciences (LP2N), IOGS-Universit\'e de Bordeaux-CNRS, 33400 Talence, France}
\author{Emmanouil E. Kriezis}
    \affiliation{School of Electrical and Computer Engineering, Aristotle University of Thessaloniki (AUTH), GR-54124 Thessaloniki, Greece}
\author{Odysseas Tsilipakos}
\email{otsilipakos@eie.gr}
    \affiliation{Theoretical and Physical Chemistry Institute, National Hellenic Research Foundation, GR-11635 Athens, Greece}

\date{\today}% It is always \today, today,

\begin{abstract}
The novel combination of a strongly-resonant optical metasurface with the MoS\textsubscript{2}/WSe\textsubscript{2} hetero-bilayer
is proposed for efficient free-space lasing enabled by the enhanced coupling between the optical and matter (exciton) states. The metasurface comprises silicon-rich nitride meta-atoms periodically arrayed in a subdiffractive lattice and overlaid with MoS\textsubscript{2}/WSe\textsubscript{2}, which provides optically-pumped gain around 1130~nm. Light emission is enabled by exploiting a quasi-bound state in the continuum in the form of a perturbed vertical magnetic dipole resonance. Following a meticulous design process guided by full-wave simulations and multipole expansion analysis, an ultralow lasing threshold of $\sim 6~\mathrm{kW/cm^2}$ is achieved. Moreover, the thermal stability of the lasing structure is examined through heat transfer simulations; stable operation with pump power densities up to a few MW/cm$^2$ (three orders of magnitude above the threshold) is predicted. These results demonstrate that MoS\textsubscript{2}/WSe\textsubscript{2}-based metasurface lasers can exhibit robust operation, paving the way for highly-performing ultrathin light-emitting surfaces. The lasing response is rigorously assessed through a highly-efficient temporal coupled-mode theory framework, verified by time-domain FEM simulations showing excellent agreement. Thus, an efficient and accurate approach to design and study metasurface lasers with arbitrary geometries and surface or bulk gain media is introduced, exhibiting significant advantages over cumbersome full-wave simulations.

\end{abstract}


\maketitle

\section{Introduction} \label{sec:Intro}

Light-emitting surfaces, where a strongly-resonant metasurface (MS) acts as the optical cavity, have concentrated significant attention during the last decade \cite{Vaskin:2019}. In particular, dielectric optical metasurfaces supporting bound states in the continuum (BICs) offer an attractive route to achieving efficient ultrathin lasing elements with tunable emission properties \cite{Vaskin:2019, Zografopoulos:2023}. Being decoupled from radiation fields, BICs are characterized by theoretically infinite radiative quality factors; their total quality factor is limited only by material absorption and fabrication imperfections. 
In practical light-emitting MSs, radiation leakage has to be allowed in order to outcouple the emitted light \cite{Cui:2018}. To this end, BICs are transformed into quasi-BICs (qBICs) with finite, yet very high, radiative quality factors. In the most common case of symmetry-protected BICs, this transformation can be performed controllably by introducing structural asymmetries in the metasurface unit cell \cite{Cui:2018, Campione:2016}. Thus, a number of MS lasers have been recently proposed and realized exploiting a range of different symmetry-protected BIC resonances. These include vertical electric \cite{Ha:2018}, magnetic \cite{Wu:2020}, and toroidal \cite{Cui:2019} dipoles, as well as higher-order dark Mie multipoles, such as electric octupoles \cite{Prokhorov:2023}, and dark surface lattice resonances \cite{Yang:2021}. 

Alongside a strongly-resonant optical state, an efficient and robust gain medium is required. The advent of two-dimensional (2D) direct-bandgap semiconductors, such as transition metal dichalcogenides (TMDs) \cite{Psilodimitrakopoulos2018}, has opened new prospects due to their extraordinary luminescence properties \cite{Elbana:2024,Cadore2024}. Recently, lasing metasurfaces where the gain is provided by TMD \textit{monolayers} have been proposed and realized \cite{Prokhorov:2023, Barth:2024}. Even more promising are TMD \textit{hetero-bilayers} forming type-II band alignment heterostructures that host bright interlayer excitons \cite{Jiang:2021}. The latter are characterized by long metastable lifetime exceeding 1~ns \cite{Palummo:2015} (in stark contrast to the ps lifetimes of intralayer excitons \cite{Palummo:2015, Selig:2018}), high quantum yield, and electrically-tunable luminescence properties in the near infrared (NIR) \cite{Lin:2024}. 

\begin{figure*}[tb]
    \centering
    \includegraphics[keepaspectratio=true]{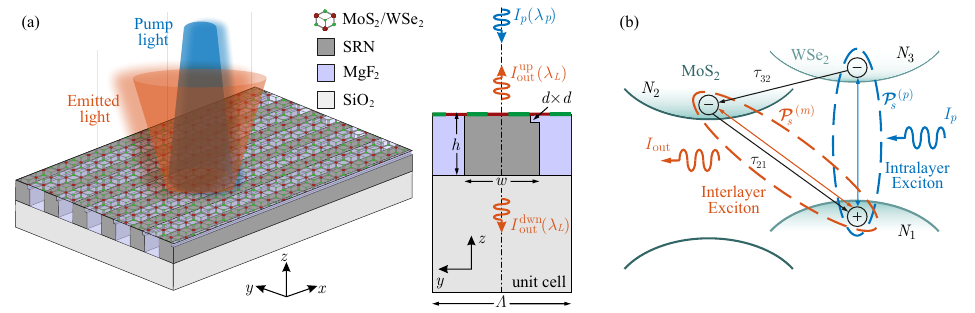}
    \caption{ (a)~Illustration of the considered metasurface laser consisting of symmetry-broken SRN stripes periodically arrayed on a SiO\textsubscript{2} substrate. The structure is planarized with a low-index cladding and is overlaid with the MoS\textsubscript{2}/WSe\textsubscript{2} hetero-bilayer. The metasurface laser is optically pumped by a normally-incident plane wave at $\lambda_p$ and emits vertically in both $\pm\hat{\mathbf{z}}$ directions at $\lambda_L$. The geometric parameters are annotated in the unit cell cross-section (right part). (b)~Energy diagram of MoS\textsubscript{2}/WSe\textsubscript{2} showing the lasing process due to the radiative recombination of interlayer excitons.}
    \label{fig:PhysStr}
\end{figure*}

One of the most promising bilayers is MoS\textsubscript{2}/WSe\textsubscript{2} that emits in the range $1100$-$1300~\mathrm{nm}$. It has been utilized in integrated nanophotonic light sources \cite{Li:2019, Huang:2025, Nousios:2024}; however, thus far it has not been leveraged for light-emitting metasurfaces. In this work, we investigate whether the combination of MoS\textsubscript{2}/WSe\textsubscript{2} with a strongly-resonant qBIC metasurface can enable efficient free-space lasing and allow for robust and stable operation. Specifically, we propose and theoretically study a dielectric MS laser composed of silicon-rich nitride (SRN) meta-atoms arrayed in a subdiffractive lattice and residing on a SiO\textsubscript{2} substrate. The metasurface is overlaid with MoS\textsubscript{2}/WSe\textsubscript{2} providing gain around $1130~\mathrm{nm}$ after being optically pumped at 740~nm. Lasing is enabled by a vertical magnetic dipole mode supported by the metasurface, i.e., a symmetry-protected BIC \cite{Murai:2020}. The optical properties of the magnetic dipole qBIC are thoroughly studied through eigenfrequency finite element method (FEM) simulations and multipolar expansion analysis \cite{Savinov:2014}. 
Upon appropriately breaking the symmetry, emitted light is radiated in the normal direction by the large coherent aperture.
The proposed MS laser is characterized by a  low lasing threshold of $\sim6~\mathrm{kW/cm^2}$, enabled by the enhanced coupling between the optical and matter (exciton) states, and an efficiency of $\sim3.7\%$. Heat transfer simulations indicate that the proposed laser can stably operate for pump irradiances up to few $\mathrm{MW/cm^2}$ (three orders of magnitude above the threshold) unhindered by deleterious thermal effects. Therefore, our results demonstrate that MoS\textsubscript{2}/WSe\textsubscript{2}-based MS lasers can exhibit efficient and robust operation, suitable for practical applications. To assess the lasing response, we utilize a  highly-efficient temporal coupled-mode theory (CMT) framework for nanophotonic lasers, rigorously implementing a semiclassical description of the light-matter interactions at the nanoscale \cite{Nousios:2023, Christopoulos:JAP2024}. The CMT results are systematically verified against full-wave nonlinear time-domain FEM (TD-FEM) simulations, showing excellent agreement. Thus, another important contribution of this work is the introduction of an efficient and accurate approach to design and study MS lasers with arbitrary geometries and surface or bulk gain media without relying on cumbersome full-wave simulations.

The rest of the paper is organized as follows: In Section~\ref{sec:Passive} the passive qBIC-metasurface is designed, while in Section~\ref{sec:Lasing} the lasing response is thoroughly assessed. Section~\ref{sec:Thermal} addresses the thermal stability of the proposed lasing element, and Section~\ref{sec:Conclusion} offers concluding remarks.

\section{Design of the quasi-dark metasurface} \label{sec:Passive}

The metasurface under study is schematically depicted in Fig.~\ref{fig:PhysStr}(a). SRN stripes are periodically arrayed in a subdiffractive lattice of pitch $\Lambda=500~\mathrm{nm}$. We opt for the ultra silicon-rich SRN compound of Ref.~\cite{Wang:2015}, characterized by a comparably high refractive index to silicon and an extended transparency window down to 605~nm. Taking advantage of the low dispersion of SRN in the NIR, we assume a wavelength-independent index of $n_\mathrm{SRN}\approx3.15$. The height of the SRN stripes is set to the typical value of $h=220~\mathrm{nm}$; their width is adjusted to $w=272~\mathrm{nm}$ for the resonance wavelength of the lasing mode, i.e., the out-of-plane magnetic dipole, to fall near the maximum of the emission spectrum of MoS\textsubscript{2}/WSe\textsubscript{2}, i.e., $1128~\mathrm{nm}$ (1.1~eV) \cite{Li:2019}. The MS resides on a SiO\textsubscript{2} substrate with $n_\mathrm{SiO_2}=1.45$. The structure is subdiffractive even in transmission for wavelengths $\lambda>n_\mathrm{SiO_2}\Lambda=725~\mathrm{nm}$ under normal incidence. The periodic configuration is planarized with a MgF\textsubscript{2} cladding so that a flat surface is available for the TMD hetero-bilayer to reside on. MgF\textsubscript{2} is a favorable cladding material characterized by both low refractive index $n_\mathrm{MgF_2}=1.37$ (resulting in  strong index contrast with SRN) and high thermal conductivity (see Section~\ref{sec:Thermal}). 
All the utilized dielectric materials are practically transparent at the considered wavelengths and are considered lossless.

\begin{figure}[tb]
    \centering
    \includegraphics[keepaspectratio=true]{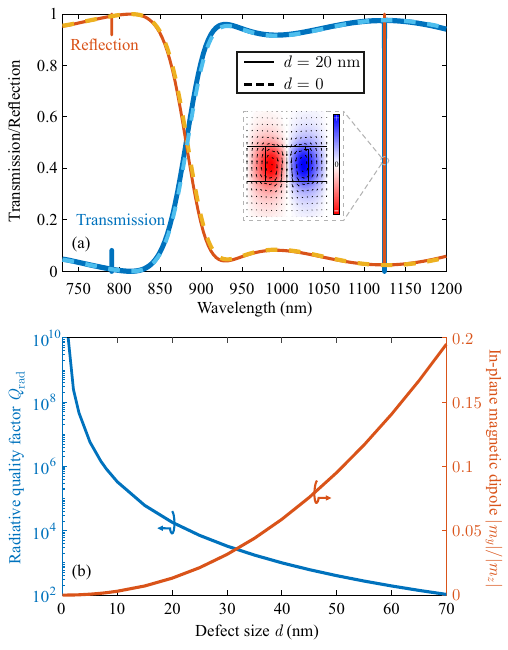}
    \caption{(a)~Transmission and reflection spectra of the examined metasurface under vertical plane-wave illumination when $d=20~\mathrm{nm}$ (solid lines) and $d=0$ (dashed lines). Inset: Field profile of the qBIC excited near $\sim1128$~nm when $d=20~\mathrm{nm}$. Color denotes $\mathrm{Re}\{ E_x\}$, whereas  arrows the magnetic field. (b)~$\mathrm{MD}_z$ mode: Radiative quality factor $Q_\mathrm{rad}$ and magnitude of the residual in-plane magnetic dipole moment $m_y$ (normalized to the dominant vertical component) as a function of the defect size $d$.}
    \label{fig:PassiveCavity}
\end{figure}

Among the dark Mie resonances supported by the dielectric metasurface, the out-of-plane magnetic dipole resonance ($\mathrm{MD}_z$) is quite favorable for lasing applications involving gain provided by 2D semiconductors  which reside atop the metasurface. This is because $\mathrm{MD}_z$ is characterized by an in-plane loop of polarization current with the electric field being tangential and interacting strongly with the sheet material. On the contrary, out-of-plane electric and toroidal dipoles have a dominant vertical electric field component, rendering them ill-suited for enhanced light-matter interaction with 2D media. We focus on the TE polarization, $\mathbf{E}=E_x\hat{\mathbf{x}}$, and examine the dark $\mathrm{MD}_z$ mode hosted by the examined passive structure (absence of MoS\textsubscript{2}/WSe\textsubscript{2}) via conducting eigenfrequency and plane-wave scattering simulations on the unit cell of the metasurface using COMSOL Multiphysics\textsuperscript{\textregistered}. The mode profile is depicted in the inset of Fig.~\ref{fig:PassiveCavity}(a), where color denotes $\mathrm{Re}\{E_x\}$ and arrows the magnetic field which lies exclusively on the $yz$ plane. When arrayed in a subdiffractive lattice, the $\mathrm{MD}_z$ mode constitutes a symmetry-protected BIC, which is completely decoupled from the radiation channel in the normal direction \cite{Murai:2020}; this is evident from the antisymmetric nature of the mode with respect to the $xz$ center plane. Hence, a square sector with side $d$ is removed from the upper right corner of the SRN stripes [Fig.~\ref{fig:PhysStr}(a)], introducing the  structural asymmetry (defect) that transforms the $\mathrm{MD}_z$ mode into a qBIC which can couple with free space. This is evident in Fig.~\ref{fig:PassiveCavity}(a), where the transmission and reflection spectra under vertical plane-wave illumination are presented for $d=0$ and $d=20~\mathrm{nm}$. Only in the latter case a sharp dip(peak) in transmission(reflection) emerges at the resonance wavelength of the $\mathrm{MD}_z$ mode, indicating that the qBIC mode can be excited. The structure also supports a magnetic quadrupole qBIC at $\sim 790~\mathrm{nm}$.

The size $d$ of the defect determines the radiative damping of $\mathrm{MD}_z$. The radiative quality factor, $Q_\mathrm{rad}$, of the mode as a function of $d$ is presented in Fig.~\ref{fig:PassiveCavity}(b) (blue line). When $d=0$ (symmetric structure), $Q_\mathrm{rad}$ diverges to infinity, whereas for $d\neq 0$ $Q_\mathrm{rad}$ becomes finite and can be tuned at will. Note that in practical realizations, the total $Q$ is bounded by material losses and fabrication-related imperfections, as well as the finite size of the MS \cite{Nousios:2025}. 
For lasing applications, values of the defect size $d$ between 15 and 40~nm (corresponding to $Q_\mathrm{rad}$ values between $\sim60\,000$ and $1\,000$) attract the main interest as they can strike a compromise between achieving a high $Q$ for low lasing threshold while permitting a sufficient amount of optical power to be emitted from the metasurface. The physical mechanism behind the transformation of the $\mathrm{MD}_z$ mode from BIC to qBIC can be revealed by conducting a multipole expansion on the polarization currents of the eigenvector, using the expressions of Ref.~\cite{Savinov:2014}. When $d=0$, the mode is solely characterized by the vertical magnetic dipole moment, $m_z$, while the other moments are zero. As $d$ increases, the mode also acquires a small in-plane/tangential magnetic dipole moment $m_y$ which increases with $d$ [red line in Fig.~\ref{fig:PassiveCavity}(b)]. In contrast to $m_z$, the $m_y$ moment is bright meaning that it scatters in the normal $\pm\hat{\mathbf{z}}$ direction. Thus, the coupling of the $\mathrm{MD}_z$ mode with the radiation fields is mediated through the residual in-plane magnetic dipole moment that the mode acquires with the introduction of the defect.  

\section{Assesment of the lasing performance} \label{sec:Lasing}

We now proceed to assess the lasing performance of the qBIC-metasurface overlaid with a  MoS\textsubscript{2}/WSe\textsubscript{2} heterostructure. The bilayer under study constitutes a three-level gain medium [Fig.~\ref{fig:PhysStr}(b)]. The pump wave excites the A intralayer excitons of WSe\textsubscript{2}, with the intralayer excitonic (pump) transition being described by the induced surface electric polarization field $\bm{\mathcal{P}}_s^{(p)}(\mathbf{r},t)$. Subsequently, the photogenerated electrons transfer from the conduction band of WSe\textsubscript{2} to that of MoS\textsubscript{2} (type-II band alignment) with characteristic carrier lifetime $\tau_{32}$ \cite{Jiang:2021}. Consequently, interlayer excitons are formed, characterized by the overall recombination (radiative and nonradiative) carrier lifetime $\tau_{21}$, for which we have $\tau_{21}\gg\tau_{32}$ resulting in the ability to attain population inversion between levels 2 (conduction band of MoS\textsubscript{2}) and 1 (valence band of WSe\textsubscript{2}). The interlayer excitonic (lasing) transition is described by a separate surface electric polarization field $\bm{\mathcal{P}}_s^{(m)}(\mathbf{r},t)$. Both of these fields follow Lorentzian oscillator equations \cite{Fietz:2012, Nousios:2024}:
\begin{equation}
    \frac{\partial^2\bm{\mathcal{P}}_s^{(k)}}{\partial t^2}+\Gamma_{iq}\frac{\partial\bm{\mathcal{P}}_s^{(k)}}{\partial t} + \omega_{iq}^2\bm{\mathcal{P}}_s^{(k)} = -\sigma_{iq}\Delta N_{iq}\bm{\mathcal{E}}_{||}, \label{eq:Pol}
\end{equation}
where $k=\{p,\,m\}$, whereas $iq=\{31,\,21\}$ denotes the levels of the gain medium related to each transition and $\Delta N_{iq}(\mathbf{r},t)=N_i(\mathbf{r},t)-N_q(\mathbf{r},t)$ is the corresponding population inversion. Furthermore, $\omega_{iq}$ and $\Gamma_{iq}$ are the central frequency and linewidth of each transition, respectively, while $\sigma_{iq}$ is the coupling strength of the transition. $\bm{\mathcal{E}}_{||}$ denotes the tangential total electric field to the TMD hetero-bilayer. The evolution of the surface carrier densities, $N_i(\mathbf{r},t)$, in each level $i=\{1,\,2,\,3\}$  of the gain medium is given by typical semiclassical carrier rate equations reading \cite{Fietz:2012, Nousios:2024}
\begin{subequations}
    \begin{align}
        \frac{\partial N_3}{\partial t} &= \frac{1}{\hbar\omega_{31}}\bm{\mathcal{E}_{||}}\cdot\frac{\partial \bm{\mathcal{P}}_s^{(p)}}{\partial t}-\frac{N_3}{\tau_{32}}, \label{eq:N_3} \\
        \frac{\partial N_2}{\partial t} &= \frac{1}{\hbar\omega_{21}}\bm{\mathcal{E}_{||}}\cdot\frac{\partial \bm{\mathcal{P}}_s^{(m)}}{\partial t}+\frac{N_3}{\tau_{32}}-\frac{N_2}{\tau_{21}}, \label{eq:N_2} \\
        \frac{\partial N_1}{\partial t} &= -\frac{1}{\hbar\omega_{31}}\bm{\mathcal{E}_{||}}\cdot\frac{\partial \bm{\mathcal{P}}_s^{(p)}}{\partial t}-\frac{1}{\hbar\omega_{21}}\bm{\mathcal{E}_{||}}\cdot\frac{\partial \bm{\mathcal{P}}_s^{(m)}}{\partial t}+\frac{N_2}{\tau_{21}}, \label{eq:N_1}
    \end{align}\label{eq:N}\end{subequations}
while the total carrier density is conserved, i.e., $N_1+N_2+N_3
\equiv N_\mathrm{tot}=\mathrm{const}$. The parameters for MoS\textsubscript{2}/WSe\textsubscript{2} are taken from Ref.~\cite{Nousios:2024} and  are typical for the examined TMD hetero-bilayer \cite{Palummo:2015, Selig:2018, Li:2019, Jiang:2021}. More specifically, $\lambda_{21}=1128~\mathrm{nm}$, $\lambda_{31}=740~\mathrm{nm}$, $\Gamma_{21}=20~\mathrm{Trad/s}$, $\Gamma_{31}=70~\mathrm{Trad/s}$, $\sigma_{21}=1.43\times10^{-8}~\mathrm{C^2/kg}$, $\sigma_{31}=5.32\times10^{-7}~\mathrm{C^2/kg}$, $\tau_{21}=2.5~\mathrm{ns}$, $\tau_{32}=100~\mathrm{fs}$, and $N_\mathrm{tot}=10^{13}~\mathrm{cm^{-2}}$. Note that recent studies have shown that $\lambda_{21}$ (the central wavelength of the interlayer excitonic transition in MoS\textsubscript{2}/WSe\textsubscript{2}) depends on the twist angle of the two constituent monolayers and for a perfectly aligned bilayer ($\mathrm{twist\mbox{-}angle}\approx 0^\circ$) $\lambda_{21}$ is closer to $1300~\mathrm{nm}$ \cite{Lin:2024}. The adopted value of $\lambda_{21}=1128~\mathrm{nm}$ corresponds to a small misalignment of a few degrees. In addition, a separate surface electric polarization field $\bm{\mathcal{P}}_s^{(\mathrm{bg})}(\mathbf{r},t)$ is used to describe the background (nonresonant) dispersive dielectric properties of MoS\textsubscript{2}/WSe\textsubscript{2}. It follows a typical single-pole Drude-Lorentz relation  \cite{Christopoulos:PRB2024}
\begin{equation}
    \frac{\partial^2\bm{\mathcal{P}}_s^{(\mathrm{bg})}}{\partial t^2}+\omega_{\mathrm{bg},0}^2\bm{\mathcal{P}}_s^{(\mathrm{bg})} = \varepsilon_0d_\mathrm{TMD}\omega_{\mathrm{bg},p}^2\bm{\mathcal{E}}_{||}, \label{eq:PolBg}
\end{equation}
where $\omega_{\mathrm{bg},0}=1.43\times10^{16}~\mathrm{rad/s}$ and $\omega_{\mathrm{bg},p}=5.99\times10^{16}~\mathrm{rad/s}$ are the background resonance and plasma frequencies fitted to the experimental data of Refs.~\cite{Liu:2014, Hsu:2019}, while $d_\mathrm{TMD}\approx1.4~\mathrm{nm}$ is the thickness of the TMD bilayer. 

Equations~\eqref{eq:Pol}, \eqref{eq:N}, and \eqref{eq:PolBg} have to be solved in conjunction with Maxwell's equations to assess the response of the proposed metasurface laser. This task involves a heavy computational burden. Thus, we resort to a much more efficient, yet rigorous and very accurate, approach based on temporal CMT that was very recently developed and demonstrated in Ref.~\cite{Nousios:2023}. It has been built upon the aforementioned semiclassical light-matter interactions using first-order perturbation theory and a limited number of approximations widely used in lasing theory. In the context of CMT, the cavity is treated as being lumped, thus the spatial dimension is removed  and the response can be obtained by a set of coupled first-order ordinary differential equations (ODEs) \cite{Christopoulos:JAP2024}. Importantly, the coefficients of the CMT equations derive directly from the physical system and are evaluated once through linear eigenvalue simulations of the unit cell of the MS. The ODE for the cavity amplitude of the lasing mode $a(t)$, normalized so that $|a(t)|^2$ is equal to the modal energy per unit length of the SRN stripes, reads \cite{Nousios:2023}
\begin{equation}
    \frac{\mathrm{d}a}{\mathrm{d}t} = -j(\omega_\mathrm{ref}-\omega_{c,0})a-\frac{1}{\tau_\mathrm{rad}}a-\xi_g\left[j\omega_\mathrm{ref}p +\frac{\mathrm{d}p}{\mathrm{d}t}\right], \label{eq:CMT_a}
\end{equation}
where $\omega_{c,0}$ is the resonance frequency of $\mathrm{MD}_z$ in the passive (cold) metasurface and $\omega_\mathrm{ref}$ is an arbitrarily chosen reference frequency that is used in place of the \textit{a-priori unknown} lasing frequency, $\omega_L$. Importantly, the actual lasing frequency is left undetermined and can be evaluated either through a Fourier transform of $a(t)$ in a post-processing step or through a closed-form relation presented in Ref.~\cite{Nousios:2023}. In addition, $\tau_\mathrm{rad}$ is the radiative cavity lifetime related to the respective quality factor through $Q_\mathrm{rad}=\omega_{c,0}\tau_\mathrm{rad}/2$, and $\xi_g$ is an overlap factor expressing the strength of interaction between the lasing mode and the sheet gain medium. Eq.~\eqref{eq:CMT_a} is complemented by the corresponding ODEs for the polarization amplitude $p(t)$ of the lasing transition and the spatially-averaged carrier rate equations. In order to validate the employed CMT framework, we also conduct nonlinear TD-FEM simulations after appropriately implementing the semiclassical description of light-matter interactions [Eqs.~\eqref{eq:Pol}, \eqref{eq:N}, and \eqref{eq:PolBg} together with Maxwell's equations] in COMSOL Multiphysics\textsuperscript{\textregistered}.   

\begin{figure}[tb]
    \centering
    \includegraphics[keepaspectratio=true]{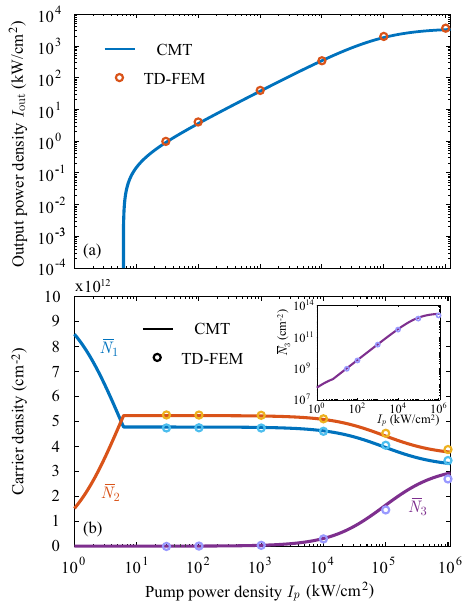}
    \caption{(a)~Total emitted power density $I_\mathrm{out}$, and (b)~spatially-averaged surface carrier densities $\bar N_i$ in the three levels of the gain medium as a function of the pump power density $I_p$, obtained by both CMT (solid lines) and TD-FEM (circular markers). Inset in (b) Carrier density $\bar N_3$ versus $I_p$ in logarithmic scale.}
    \label{fig:LasingCW}
\end{figure}

\begin{figure*}[tb]
    \centering
    \includegraphics[keepaspectratio=true]{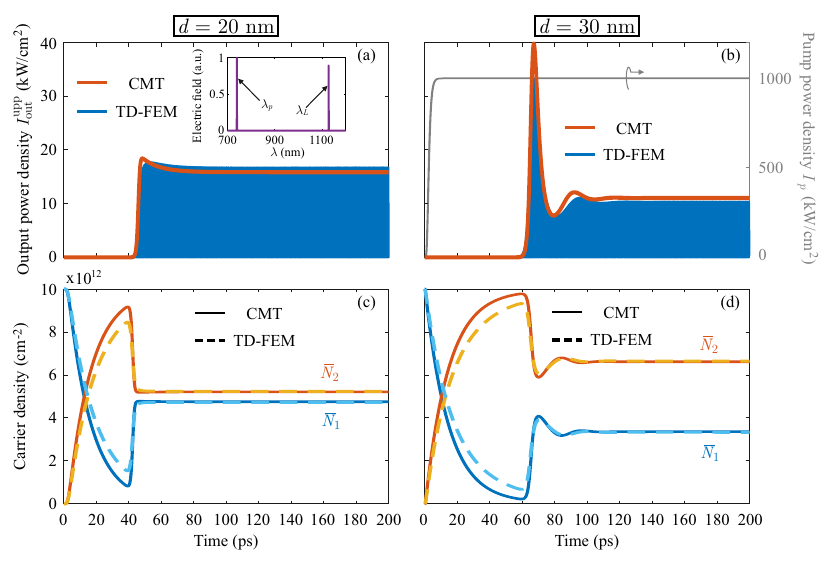}
    \caption{Temporal evolution of (a,\,b)~the output power density in air, $I_\mathrm{out}^\mathrm{up}$, and (c,\,d)~the spatially-averaged surface carrier densities $\bar N_1$ and $\bar N_2$ evaluated through both CMT and TD-FEM. The pump power density is smoothly switched-on until reaching the constant value of $I_p=1~\mathrm{MW/cm^2}$ [gray curve in (b)]. In (a,\,c) the defect size is $d=20~\mathrm{nm}$, while in (b,\,d) $d=30~\mathrm{nm}$. Inset in (a) normalized optical spectrum of the electric field in air comprising the reflected pump wave at $\lambda_p$ and the upwards emitted wave at $\lambda_L$.} %assessed through TD-FEM and
    \label{fig:LasingDyn}
\end{figure*}

We initially set the defect size to $d=20~\mathrm{nm}$ for which we have $\lambda_{c,0} = 1127.76~\mathrm{nm}\approx\lambda_{21}$ and $Q_\mathrm{rad} = 17\,633$. The MS laser is then pumped by a normally-incident plane wave with pump power density $I_p$ at the pump wavelength of $\lambda_p=\lambda_{31}=740~\mathrm{nm}$ and emits vertically in both $\pm\hat{\mathbf{z}}$ directions with a total power density $I_\mathrm{out}$ at the lasing wavelength $\lambda_L$; the upward(downward) output power density in air(substrate) is denoted as $I_\mathrm{out}^\mathrm{up}$($I_\mathrm{out}^\mathrm{dwn}$). The light-light curve ($I_\mathrm{out}$ versus $I_p$) evaluated through both CMT (solid line) and TD-FEM (circular markers) is depicted in Fig.~\ref{fig:LasingCW}(a). A broad range of pump power density levels is covered ($1~\mathrm{kW/cm^2}$ to $1~\mathrm{GW/cm^2}$) and the agreement between the two methods is excellent. Note that in CMT the total output power density is calculated through  $I_\mathrm{out}=(2/\tau_\mathrm{rad})|a|^2/\Lambda$. In TD-FEM, on the other hand, $I_\mathrm{out}$ is assessed by recording the electric field at an appropriate fictitious plane above/below the metasurface unit cell, followed by a post-processing Fourier transform to separate the reflected/transmitted pump wave from the emitted wave in the respective direction [inset of Fig.~\ref{fig:LasingDyn}(a)]. 

The examined metasurface laser exhibits a very low lasing threshold of $I_{p,\mathrm{th}}=6.16~\mathrm{kW/cm^2}$ and a total efficiency ($I_\mathrm{out}/I_p$) of $\sim3.7\%$. This is due to the efficient coupling between the interlayer excitons of MoS\textsubscript{2}/WSe\textsubscript{2} and the magnetic dipole qBIC mode of the metasurface, as well as the enhanced absorption of the pump light by the intralayer excitons of the WSe\textsubscript{2} constituent monolayer. Although the incident pump wave is not coupled to a high quality-factor mode of the MS, the absorption efficiency, $\eta_p$, is still significant with $\eta_p=9.56\%$. The power emitted in air (upwards emission) is 42.3\% of the total emitted power, while the rest (57.7\%) is emitted in the substrate (downwards emission), as evaluated through nonlinear TD-FEM simulations or linear eigenvalue simulations. Note that for a free-standing structure (i.e., absence of substrate), the emitted power would have been equally split in the two $\pm\hat{\mathbf{z}}$ directions, since the lasing mode radiates in the far-field as an in-plane magnetic dipole with no other multipole contributing. In addition, CMT and TD-FEM agree in the prediction of the lasing frequency, for which it holds $\lambda_L\sim\lambda_{21}$ since we had initially aligned the cavity resonance with the peak of the gain spectrum, i.e., $\lambda_{c,0}\approx\lambda_{21}$. We stress that in contrast to other models our CMT framework does \textit{not} require $\lambda_{c,0}\equiv\lambda_{21}$ and the detuning between them can be selected freely. The agreement between CMT and TD-FEM is further highlighted in Fig.~\ref{fig:LasingCW}(b), where the spatially-averaged surface carrier densities, $\bar N_i$, in the three levels of the gain medium are depicted as a function of $I_p$.

In Fig.~\ref{fig:LasingDyn}(a), we depict the temporal evolution of the output power density in air, $I_\mathrm{out}^\mathrm{up}$, evaluated through both CMT and TD-FEM, when $I_p$ is smoothly switched-on until reaching the constant value of $1~\mathrm{MW/cm^2}$ [gray curve in Fig.~\ref{fig:LasingDyn}(b)]. The corresponding dynamic response of $\bar N_1$ and $\bar N_2$ is presented in Fig.~\ref{fig:LasingDyn}(c). Once again, the agreement between the two methods is very good both at steady-state and during the transient. The same calculations are also performed for a greater defect size, $d=30~\mathrm{nm}$, with the corresponding results presented in Figs.~\ref{fig:LasingDyn}(b,\,d). By increasing the introduced asymmetry, the radiative quality factor decreases to $Q_\mathrm{rad}=3\,250$ [cf. Fig.~\ref{fig:PassiveCavity}(b)]. Concurrently, the resonance wavelength is blueshifted from the central emission wavelength to the value $\lambda_{c,0}=1124.5~\mathrm{nm}$, thus, leading to a suboptimal coupling between the gain medium and the lasing mode. Therefore, the lasing threshold is increased to $I_{p,\mathrm{th}}=11.13~\mathrm{kW/cm^2}$, an almost twofold increase compared to $d=20~\mathrm{nm}$. This is indirectly captured in Fig.~\ref{fig:LasingDyn}(d), where the population inversion at steady state, $\Delta\bar N_{21, \mathrm{ss}}\propto I_{p,\mathrm{th}}$ \cite{Nousios:2023}, is much larger than that of Fig.~\ref{fig:LasingDyn}(c). In addition, the output power density at steady-state is now lower for the same level of $I_p$ [cf. Figs.~\ref{fig:LasingDyn}(a,\,b)], while the transient effect is more pronounced with intense relaxation oscillations. The lasing wavelength obtained by TD-FEM simulations is $\lambda_L=1124.86~\mathrm{nm}$ in this case, in accordance with that estimated through CMT ($\lambda_L=1124.6~\mathrm{nm}$).

The importance of the results in Fig.~\ref{fig:LasingDyn} should not be overlooked. Using only first order simplifications and spatial averaging, we were able to reproduce the temporal response of a highly inhomogeneous metasurface accurately predicting not only the emitted power but also the carrier densities in the gain medium. CMT can be utilized to probe the underlying physics in complex systems, enables physical interpretations in an efficient manner, and allows for quantitative assessments of the performance.

\section{Thermal stability analysis} \label{sec:Thermal}

In Fig.~\ref{fig:LasingCW}(a), we have examined the CW lasing response of the proposed MS for an extended range of pump power density levels, up to the point where lasing is fully saturated due to the quenching of the intralayer excitonic transition (pump absorption saturation). The latter occurs for $I_p>100~\mathrm{MW/cm^2}$, i.e., four orders of magnitude above the lasing threshold. Nonetheless, in practical nanophotonic lasing structures the emission process can be hampered at much lower irradiances due to deleterious thermal effects, which can become important in the case of CW pumping (they are typically benign in the case of pulsed pumping). In fact, they are even more pronounced in compact dielectric nanostructures, where the small cavity volume and the absence of metallic interfaces limit the heat removal capabilities \cite{Wen:2022}. Thus, heat generated by Joule losses remains trapped inside the nanophotonic laser leading to a substantial increase of the lattice temperature of the gain medium and the subsequent degradation of its luminescence properties.

\begin{figure}[tb]
    \centering
    \includegraphics[keepaspectratio=true]{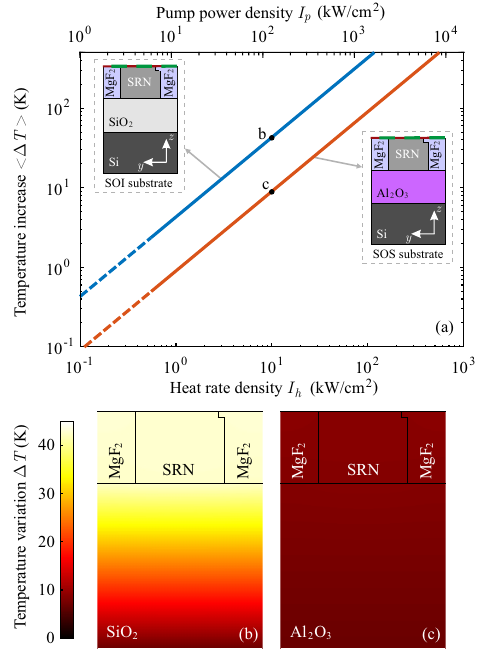}
    \caption{(a)~Temperature increase in MoS\textsubscript{2}/WSe\textsubscript{2} (spatially averaged) as a function of the generated heat rate density $I_h$ for the two different configurations shown in the insets. Dashed lines indicate operation below the lasing threshold. (b,\,c)~Spatial distribution of the temperature variation in the metasurface unit cell for $I_h=10~\mathrm{kW/cm^2}$ ($I_p \approx 170~\mathrm{kW/cm^2}$) when the metasurface resides on top of a (a)~standard silicon-on-insulator, and (b)~silicon-on-sapphire substrate.}
    \label{fig:Thermal}
\end{figure}

The heat removal capability of the examined structure under CW pumping is assessed through heat transfer simulations with COMSOL Multiphysics\textsuperscript{\textregistered}. We initially examine a practical realization of the proposed metasurface laser built on a standard silicon-on-insulator (SOI) wafer [left inset of Fig.~\ref{fig:Thermal}(a)]. The thickness of the buried oxide (SiO\textsubscript{2}) is $0.5~\mathrm{\mu m}$. For the silicon handle layer,  a thickness of $10~\mathrm{\mu m}$ is included in the simulations before truncating with a constant-temperature boundary condition. As the MS consists of transparent materials at both the pump and lasing wavelengths, the sole heat source is the fraction of the pump power absorbed by the 2D gain medium that is not offered for the amplification of the lasing mode. More specifically, the heat rate density, $I_h$, is given by $I_h=\eta_p I_p-I_\mathrm{out}$. We also assume that the heat is uniformly generated over the MoS\textsubscript{2}/WSe\textsubscript{2} layer. The simulations are conducted with the unit cell of the metasurface, using continuity boundary conditions at the lateral sides, convective boundary conditions at the interfaces with air [heat transfer coefficient equal to $h=10~\mathrm{W/(m^2K)}$ \cite{Tsilipakos:2009}], and a constant temperature, $T_0=293~\mathrm{K}$, boundary condition at the bottom side of Si. The thermal conductivities of the underlying materials are $K_\mathrm{Si}=K_\mathrm{SRN}=145~\mathrm{W/(m\,K)}$, $K_\mathrm{MgF_2}=28~\mathrm{W/(m\,K)}$, and $K_\mathrm{SiO_2}=1.4~\mathrm{W/(m\,K)}$. In Fig.~\ref{fig:Thermal}(a) (blue curve), we depict the spatially-averaged temperature increase in MoS\textsubscript{2}/WSe\textsubscript{2} compared with the ambient temperature $T_0=293~\mathrm{K}$, $\langle\Delta T\rangle$,  as a function of $I_h$ and $I_p$. The lasing response is expected to be unobstructed by deleterious thermal effects for almost two orders of magnitude above the lasing threshold, where $\langle\Delta T\rangle$ is well below 100~K. 

As $I_p$ approaches $1~\mathrm{MW/cm^2}$, the lattice temperature increases excessively. The low thermal conductivity of the SiO\textsubscript{2} substrate is the main reason for the excessive temperature increase, as it restricts the generated heat from being effectively funneled into the high thermal-conductivity Si layer, where it can be dissipated. This is also evident in Fig.~\ref{fig:Thermal}(b), where the spatial variation of the temperature in the $yz$ plane is depicted when $I_h=10~\mathrm{kW/cm^2}$ ($I_p\approx 170~\mathrm{kW/cm^2}$). A significant improvement of the heat removal capability can be achieved by considering the examined metasurface laser on top of a silicon-on-sapphire (SOS) wafer [right inset of Fig.~\ref{fig:Thermal}(a)]. Sapphire (Al\textsubscript{2}O\textsubscript{3}) possesses twenty times higher thermal conductivity than  SiO\textsubscript{2} [$K_\mathrm{Al_2O_3}=28~\mathrm{W/(m\,K)}$] and a comparably low refractive index ($n_\mathrm{Al_2O_3}=1.75$). The corresponding spatially-averaged temperature increase over MoS\textsubscript{2}/WSe\textsubscript{2} is presented in Fig.~\ref{fig:Thermal}(a) (red curve). The introduction of the high thermal-conductivity substrate results in a fivefold reduction of $\langle\Delta T\rangle$ for the same level of $I_p$ [cf. Figs.~\ref{fig:Thermal}(b,\,c)], which consequently enables the operation of the metasurface laser at much higher pump power density levels up to few $\mathrm{MW/cm^2}$ (three orders of magnitude above threshold). 

Note that the presence of the high thermal-conductivity MgF\textsubscript{2} cladding also contributes to heat removal in both of the investigated configurations. This is because the MgF\textsubscript{2} cladding and the SRN core form a uniform high-thermal-conductivity layer across the entire lateral extent of the metasurface, which effectively drains the generated heat towards the substrate \cite{Wen:2022, Nousios:2023}.  

\section{Conclusion} \label{sec:Conclusion}

In summary, we have proposed a metasurface laser comprising SRN stripes periodically arrayed in a subdiffractive lattice and overlaid with the contemporary MoS\textsubscript{2}/WSe\textsubscript{2} TMD hetero-bilayer. Highly efficient lasing in the NIR is enabled by the enhanced coupling between the qBIC magnetic dipole mode supported by the metasurface and the long-lived interlayer excitons hosted by the 2D gain medium. The modal properties of the quasi-dark resonance and the physical mechanism behind its transformation from BIC to qBIC are elucidated through eigenfrequency simulations and multipole expansion analysis. The proposed structure is characterized by very low lasing threshold of only few $\mathrm{kW/cm^2}$. A comprehensive thermal stability analysis through heat transfer simulations has shown that the laser can function free from deleterious thermal effect in a wide range of pump power densities (up to three orders of magnitude above threshold), where the metasurface laser can deliver practical levels of output power. 

The lasing response of the examined metasurface has been thoroughly assessed through a highly efficient temporal CMT framework verified by full-wave nonlinear TD-FEM simulations. The agreement between the two methods is excellent both at steady-state and during the transient. The validation of the CMT framework opens new avenues for efficiently designing and studying elaborate metasurface laser structures based on 2D or 3D gain media without resorting to time-consuming full-wave nonlinear simulations \cite{Christopoulos:2025}.

\begin{acknowledgments}
OT acknowledges Dr. Thomas Koschny for fruitful discussions.
\end{acknowledgments}

\section*{Conflict of interest Statement}
The authors have no conflicts to disclose.

\section*{Data Availability Statement}
The data that support the findings of this study are available from the corresponding authors upon reasonable request.

\end{document}